\documentstyle[eqsecnum,aps,floats,preprint]{revtex}
\begin{document}
\preprint{LBL--37244}
\title{$W\gamma$ and $Z\gamma$ Production at Tevatron\footnote{
Invited talk given at the International Symposium on Vector Boson
Self--Interactions, UCLA, February 1--3, 1995.}$^,$\footnote{
This work was supported by
the Director, Office of Energy Research,
Office of High Energy and Nuclear Physics,
Division of High Energy Physics of the U.S.~Department of Energy
under Contract DE-AC03-76SF00098.}}

\author{H.~Aihara}
\address{Lawrence Berkeley Laboratory\\
Berkeley, California 94720}
\maketitle

\begin{abstract}
We present results from CDF and D\O\ on $W\gamma$ and $Z\gamma$ productions
in $p\bar{p}$ collisions at $\sqrt{s}=1.8~{\rm TeV}.$
The goal of the analyses is to test
the non-abelian self-couplings of the $W$, $Z$ and photon,
one of the most direct consequences
of the $SU(2)_L\otimes U(1)_Y$ gauge symmetry.
We present  direct measurements of $WW\gamma$ couplings  and
limits on  $ZZ\gamma$ and $Z\gamma\gamma$ couplings,
based on $p\bar{p}\rightarrow \ell\nu\gamma + X$ and
$p\bar{p}\rightarrow \ell\ell\gamma+X$ events, respectively,
observed during the 1992--1993 run of the Fermilab Tevatron Collider.
\end{abstract}

\section*{Introduction}

Direct measurement of the $WW\gamma$  gauge boson couplings is possible
through study of  $W\gamma$ production in $p\bar{p}$ collisions at
$\sqrt{s}=1.8$ TeV.
The most general effective Lagrangian~\cite{Hagiwara},
invariant under $U(1)_{EM}$, for the $WW\gamma$ interaction
contains four coupling parameters, CP--conserving $\kappa$ and $\lambda$, and
CP--violating $\tilde{\kappa}$ and $\tilde{\lambda}$.
The CP--conserving parameters are related to the magnetic dipole ($\mu_W$) and
electric quadrupole ($Q_W^e$) moments of the $W$ boson, while
the CP--violating parameters are related to the electric dipole
($d_W$) and the magnetic quadrupole ($Q_W^m$) moments:
\mbox{$\mu_W=(e/2m_W)(1+\kappa+\lambda),$}
$Q_W^e=(-e/m_W^2)(\kappa-\lambda),\
d_W=(e/2m_W)(\tilde{\kappa}+\tilde{\lambda}),\
Q_W^m=(-e/m_W^2)(\tilde{\kappa}-\tilde{\lambda})$~\cite{Kim}.
In the Standard Model (SM) the $WW\gamma$ couplings at the tree level are
uniquely determined by
the $SU(2)_L\otimes U(1)_Y$ gauge symmetry:~$\kappa=1$
$(\Delta \kappa\equiv \kappa - 1 =0)$, $\lambda=0$,
$\tilde{\kappa}=0$, $\tilde{\lambda}=0.$
The direct and  precise measurement of the $WW\gamma$ couplings
is of interest since
the existence of anomalous couplings, i.e.~measured values different from the
SM predictions,  would indicate the presence of physics beyond the SM.
A $WW\gamma$ interaction Lagrangian with constant, anomalous couplings
violates unitarity at
high energies, and, therefore,
the coupling parameters must be modified to include
form factors (e.g.~$ \Delta\kappa (\hat{s})=
\Delta\kappa/(1+\hat{s}/\Lambda_W^2)^n,$
where $\hat{s}$ is the square of the invariant mass of the $W$ and the photon,
$\Lambda_W$ is the form factor scale, and
$n=2$ for a dipole form factor)~\cite{Baur}.

The study of the $Z\gamma$ production in $p\bar{p}$ collision is also an
important test of the SM description of gauge-boson self-interactions.
Since the photon does not couple directly to the $Z$ in the SM, this study is
sensitive to anomalous couplings beyond the SM.
The most general $ZZ\gamma$ ($Z\gamma\gamma$) vertex function is characterized
by a set of four coupling parameters $h_{1-4}^{Z(\gamma)}$~\cite{Hagiwara}.
All these coupling parameters vanish at tree level within the framework
of the SM.
The couplings $h_3^V$ and $h_4^V$ conserve CP, while $h_1^V$ and $h_2^V$
are CP-violating.
Similarly to the $WW\gamma$ anomalous couplings,
the $ZZ\gamma(Z\gamma\gamma)$ couplings must be regulated by
generalized dipole form factors:
$(h_i^V(\hat{s})=h_{i0}^V/(1+\hat{s}/\Lambda_Z^2)^n,$
where $h_{i0}^V$ represents the low energy ($\hat{s}=0$) limit for the
couplings, and $n=3$ for $h_{1,3}^V$ and $n=4$ for $h_{2,4}^V.$
Here the values for $n$ were chosen so that the unitarity is preserved and that
all terms in the matrix element proportional to $h_{i0}^V$ have the
same asymptotic energy behavior.
At the Tevatron, the $W\gamma$ production is insensitive to
the form factor effects for $\Lambda_W>$ a few 100 GeV, whereas the form factor
effects
cannot be ignored for $Z\gamma$ production
due to the higher power
of $\hat{s}$ dependence in the $ZZ\gamma (Z\gamma\gamma)$ vertex function.

We present studies of the $WW\gamma$ and $ZZ\gamma(Z\gamma\gamma)$ couplings
based on  $p\bar{p}\rightarrow \ell\nu\gamma+X$  and
$p\bar{p}\rightarrow \ell\ell\gamma$ $(\ell=e,\mu)$ events
observed with the CDF~\cite{CDF} and D\O\ detector~\cite{DO}
during the 1992--1993 run of the Fermilab Tevatron Collider,
corresponding to integrated luminosities of $\sim 20\ {\rm pb}^{-1}$
for CDF and $\sim 14\ {\rm pb}^{-1}$ for D\O.
The $\ell\nu\gamma$ events contain
the $W\gamma$ production process, $p\bar{p}\rightarrow W\gamma+X$ followed by
$W\rightarrow \ell\nu$, and
the radiative $W\rightarrow \ell\nu\gamma$ decay where the photon
originates from bremsstrahlung of the charged lepton.
Anomalous coupling parameters enhance the $W\gamma$ production
with a large $\hat{s}$, and thereby result in
an excess of events  with
high transverse energy, $E_T$, photons, well separated from the
charged lepton.
The $\ell\ell\gamma$ events contain the radiative
$Z\rightarrow \ell\ell\gamma$ decay,
the direct $Z\gamma$ production where the photon
is radiated from one of the annihilating quarks, and the possible
$Z\gamma$ events due to the anomalous $Z$-$\gamma$ couplings.
The presence of the $Z$-$\gamma$ couplings will also be signaled by
an excess of $Z$ production with high $E_T$ photons.

\section*{Photon Detection at CDF and D\O}

Since the good detection of the photon is the key to the $W\gamma$ and
$Z\gamma$
measurements, we briefly review how photons are detected by the CDF  and
D\O\
detectors.
A photon is identified as a calorimeter energy cluster satisfying the
following condition.
A calorimeter cluster must (i) have a high electromagnetic energy fraction;
(ii) be isolated;
(iii) have shower shape consistent with a single photon; and
(iv) have no tracks pointing to it.
Table 1 summarizes the actual conditions required by CDF and D\O.

To test  shower shape of the cluster
CDF uses the central electromagnetic strip chambers~\cite{CES}
(CES) placed after $\sim 6.3$
radiation lengths in the central electromagnetic calorimeter.
The CES determines shower position and transverse
development of an electromagnetic shower at shower maximum by
measurement of the charge deposition on orthogonal, fine-grained (1.5~cm
spacing)  strips and wires.
D\O\ tests both longitudinal and transverse shower shapes
including correlations
between energy deposits in the fine-grained calorimeter cells~\cite{MEENA}.
The D\O\ electromagnetic calorimeter module has 4 longitudinal layers.
Each of layers 1, 2 and 4 is segmented transversely
to $\Delta\eta\times \Delta\phi=0.1\times 0.1$,
while the  third layer, which typically contains $65\%$ of
the EM energy, has segmentation of
$\Delta\eta\times \Delta\phi=0.05\times 0.05.$
($\eta$ is the pseudorapidity defined as $\eta=-\ln (\tan(\theta/2))$, $\theta$
being the polar angle with repect to the beam axis.
$\phi$ is the azimuthal angle.)

Both CDF and D\O\ found that the detection efficiency for photons depends on
$E_T^\gamma$  due to the isolation requirement.
D\O\ found its cluster shape requirement also results in the $E_T$ dependence.
The overall photon detection efficiency was obtained by combining
this $E_T$--dependent efficiency with the probabilities of losing a photon
due to $e^+e^-$ pair conversions and due to an overlap with a random track
in the event.
Table 2 summarizes the photon detection efficiencies at CDF and D\O.

\begin{table}
\caption{Summary of photon detection at CDF and D\O}
\label{table1}
\begin{tabular}{lll}
  & CDF & D\O \\
\tableline
  &   & \\
detection & $|\eta|<1.1$  & $|\eta|<1.1$ \\
region    & ($1.1<|\eta|<2.4$~\tablenote{Analysis in progress.})
& $1.5<|\eta|<2.5$ \\
 & & \\
minimum $E_T^\gamma$ & 7~GeV & 10~GeV \\
  & & \\
EM fraction & $HAD/EM$ & $EM/Total>0.9$ \\
            & $<0.055+0.00045\times E({\rm GeV})$\ \ \ \ \ \ \  &  \\
   & & \\
Isolation & $(E_T(0.4)-E_T^\gamma)/E_T^\gamma<0.15$~\tablenote{
$E_T(0.4)$ is the $E_T$ in a cone of
$\Delta R=\sqrt{(\Delta \eta)^2+(\Delta\phi)^2}=0.4$ around the photon
candidate.
$p_T(0.4)$ is the sum of $p_T$ of the charged tracks within the same cone.} &
$(E(0.4)-EM(0.2))/EM(0.2)<0.10$~\tablenote{
$E(0.4)$ is the total energy inside a cone of radius $\Delta R=0.4,$
and $EM(0.2)$ is the EM energy inside a cone of 0.2.} \\
   & $p_T(0.4)<2$~GeV/c & \\
  & & \\
Shower shape\ \ \ \ & transverse & longitudinal/transverse \\
   & & \\
No track & No matching tracks & No matching tracks\\
   & &   \\
\end{tabular}
\end{table}

\begin{table}
\caption{Summary of photon detection efficiency.}
\label{table2}
\begin{tabular}{lccc}
   & CDF & \multicolumn{2}{c}{D\O}  \\
\tableline
   & &  & \\
    & $|\eta|<1.1$  & $|\eta|<1.1$ & $1.5<|\eta|<2.5$ \\
\tableline
$E_T^\gamma>25$ GeV& $0.804\pm 0.023$ & $0.74\pm 0.07$ & $0.58\pm 0.05$ \\
\ \ \ \ \ =\ 10  &   & $0.43\pm 0.04$ & $0.38\pm 0.03$ \\
\ \ \ \ \ =\ \ 7   & $0.731\pm 0.021$ & & \\
\end{tabular}
\end{table}

\section*{$W\gamma$ analysis}

The $W\gamma$ candidates were obtained by searching for events containing
an isolated lepton ($e$ or $\mu$) with high $E_T$, large missing transverse
energy, \mbox{$\not\!\!E_T$}, and an isolated photon.
Table 3 summarizes geometrical and kinematic selection as well as
integrated luminosity used in each channel.
Both CDF and D\O\ required that the separation between a photon and
a lepton be $\Delta R_{\ell\gamma}>0.7.$
This requirement suppresses the contribution of the radiative $W$ decay
process.
The CDF observed 18 $W(e\nu)\gamma$ candidates and 7 $W(\mu\nu)\gamma$
candidates~\cite{CDF_WG}, while the D\O\ observed 11 $W(e\nu)\gamma$ candidates
and 12 $W(\mu\nu)\gamma$ candidates~\cite{DO_WG}.

\begin{table}
\caption{Summary of $W\gamma$ event selection.}
\label{table3}
\begin{tabular}{lcccc}
   & \multicolumn{2}{c}{CDF} & \multicolumn{2}{c}{D\O}  \\
\tableline
   & &  & & \\
  & \multicolumn{1}{c}{$e\nu\gamma$} & \multicolumn{1}{c}{$\mu\nu\gamma$} &
\multicolumn{1}{c}{$e\nu\gamma$} & \multicolumn{1}{c}{$\mu\nu\gamma$} \\
   & & & & \\
Geometry  & $|\eta_e|<1.1$  & $|\eta_\mu|<0.6$ &
$|\eta_e|<1.1$ & $|\eta_\mu|<1.7$ \\
  &  &   &  $1.5<|\eta_e|<2.5$ & \\
  & & & & \\
 & \multicolumn{2}{c}{$|\eta_\gamma|<1.1$} &
\multicolumn{2}{c}{$|\eta_\gamma|<1.1$,\ $1.5<|\eta_\gamma|<2.5$}\\
  & & & & \\
Kinematics & $E_T^e>20$ & $p_T^\mu>20$ & $E_T^e>25$ & $p_T^\mu>15$ \\
(in GeV)    & \mbox{$\not\!\! E_T>20$} & \mbox{$\not\!\! E_T>20$} &
 \mbox{$\not\!\! E_T>25$} & \mbox{$\not\!\! E_T>15$} \\
   & \multicolumn{2}{c}{$E_T^\gamma>7$} & \multicolumn{2}{c}{$E_T^\gamma>10$}
  \\
   & & & & \\
   & \multicolumn{2}{c}{$\Delta R_{\ell\gamma}>0.7$}
& \multicolumn{2}{c}{$\Delta R_{\ell\gamma}>0.7$}\\
  & & & & \\
$\int Ldt\ {\rm pb}^{-1}$ & $19.6\pm 0.7$ & $18.6\pm 0.7$ & $13.8\pm 0.7$ &
$13.7\pm 0.7$ \\
\end{tabular}
\end{table}

\begin{table}
\caption{Summary of $W\gamma$ data and backgrounds.}
\label{table4}
\begin{tabular}{lcccc}
   & \multicolumn{2}{c}{CDF} & \multicolumn{2}{c}{D\O}  \\
\tableline
   & &  & & \\
  & \multicolumn{1}{c}{$e\nu\gamma$} & \multicolumn{1}{c}{$\mu\nu\gamma$} &
\multicolumn{1}{c}{$e\nu\gamma$} & \multicolumn{1}{c}{$\mu\nu\gamma$} \\
Source: & & & & \\
\ \ $W+$jets & $4.6\pm 1.8$ & $1.9\pm 0.6$ & $ 1.7\pm 0.9$ & $1.3\pm 0.7$ \\
\ \ $Z\gamma$ & $0.43\pm 0.02$ & $ 1.14\pm 0.06$ & $0.11\pm 0.02$
& $2.7\pm 0.8$ \\
\ \  $W(\tau\nu)\gamma$ & $0.29\pm 0.02$ & $0.15\pm 0.01$ & $0.17\pm 0.02$ &
$0.4\pm 0.1$ \\
   & & & & \\
Total background & $5.3\pm 1.8$ & $3.2\pm 0.6$ & $ 2.0\pm 0.9$ & $4.4\pm 1.1$
\\
  & & & & \\
Data & 18 & 7 & 11 & 12 \\
 & & & & \\
Signal & $12.7\pm 4.6$ & $3.8\pm 2.7$ & $9.0^{+4.2}_{-3.1}\pm 0.9$ &
$7.6^{+4.4}_{-3.2}\pm 1.1$ \\
\end{tabular}
\end{table}

\begin{table}
\caption{Comparison of data and the SM prediction for $W\gamma$.}
\label{table5}
\begin{tabular}{lcccc}
   & \multicolumn{2}{c}{CDF} & \multicolumn{2}{c}{D\O}  \\
\tableline
   & &  & & \\
  & \multicolumn{1}{c}{$e\nu\gamma$} & \multicolumn{1}{c}{$\mu\nu\gamma$} &
\multicolumn{1}{c}{$e\nu\gamma$} & \multicolumn{1}{c}{$\mu\nu\gamma$} \\
 & & & & \\
Signal & $12.7\pm 4.6$ & $ 3.8\pm 2.7$ & $9.0^{+4.2}_{-3.1}\pm 0.9$ &
$7.6^{+4.4}_{-3.2}\pm 1.1$ \\
   & & & & \\
SM prediction & $15.4\pm 0.7$ & $7.9\pm 0.4$ & $ 6.9\pm 1.0$ & $6.7\pm 1.2$ \\
   & & & & \\
$\sigma_{W\gamma}$ ($E_T^\gamma>\ 7{\rm GeV}$,$\Delta R_{\ell\gamma}>0.7$)
pb & $141.7 \pm 53$ & $83\pm 59$  & & \\
$\sigma_{W\gamma}$ ($E_T^\gamma>10{\rm GeV}$,$\Delta R_{\ell\gamma}>0.7$)
pb\ \ \ \  &  & & $147^{+73}_{-56}$ & $127^{+78}_{-61}$ \\
  & & & & \\
$e+\mu$ combined  & \multicolumn{2}{c}{$122\pm 42$ pb} &
\multicolumn{2}{c}{$138^{+55}_{-43}$ pb}  \\
 & & & & \\
SM prediction   & \multicolumn{2}{c}{$172\pm 26$ pb} &
\multicolumn{2}{c}{$112\pm 10$ pb}  \\
\end{tabular}
\end{table}

The background estimate, summarized in Table 4, includes contributions from:
$W+$jets, where a jet is misidentified as a photon;
$Z\gamma$, where the $Z$ decays to $\ell^+\ell^-$, and one of the leptons is
undetected or is mismeasured by the detector and contributes to \mbox{$\not\!\!
E_T$};
$W\gamma$ with $W\rightarrow \tau\nu$ followed by $\tau\rightarrow
\ell\nu\bar{\nu}.$
The $W+$jets background was estimated using the probability,
${\cal P}(j\rightarrow ``\gamma")$, for a jet to be misidentified as a
photon determined as a function of $E_T$ of the jet
by measuring the fraction of jets in a sample of multijet events
that pass our photon identification requirements.
For the photon criteria used by CDF,
${\cal P}(j\rightarrow ``\gamma")\sim 8\times 10^{-4}$ at $E_T^j=9$ GeV,
decreasing exponentially to
${\cal P}(j\rightarrow ``\gamma")\sim 1\times 10^{-4}$ at $E_T^j=25$ GeV.
For the photon criteria used by D\O,
${\cal P}(j\rightarrow ``\gamma")\sim 4\times 10^{-4}~(6\times 10^{-4})$
in the central (endcap) calorimeter, and varies only slowly with $E_T^j.$
The total number of $W+$jets background events was calculated
by applying ${\cal P}(j\rightarrow ``\gamma")$
to the observed $E_T$ spectrum of jets in the inclusive $W(\ell\nu)$ sample.
The backgrounds due to $Z\gamma$ and $W\rightarrow \tau\nu$ were
estimated from Monte Carlo simulations.

The kinematic and geometrical acceptance was calculated as a function of
coupling parameters, $\Delta\kappa$ and $\lambda$, using the Monte Carlo
program of Baur and Zeppenfeld, in which
the $W\gamma$ production and radiative decay processes are generated to
leading order, and higher order QCD effects are approximated by a K-factor.
Both CDF and D\O\ used the MRSD\_' structure functions and
simulated the $p_T$ distribution of the $W\gamma$ system using the observed
$p_T$ spectrum of the $W$ in the inclusive $W(\ell\nu)$ sample.
The generated events underwent a detector simulation.
Table 5 shows the comparison between the observed signal and the SM prediction.
CDF obtained the $W\gamma$ cross section for photons with $E_T^\gamma>7$ GeV
and $\Delta R_{\ell\gamma}>0.7$ from a combined $e+\mu$ sample:
$\sigma(W\gamma)=122\pm 42$~pb,
while the SM predicts $172\pm 26$~pb.
D\O\ obtained $\sigma(W\gamma)=138^{+55}_{-43} $~pb for
photons with $E_T^\gamma>10$ GeV and $\Delta R_{\ell\gamma}>0.7$,
and the SM predicts $112\pm 10$~pb.
Here we used  BR($W\rightarrow \ell\nu)=0.108.$
The observed cross section agrees with the SM prediction within errors.

Figures 1 and 2 show that data and the SM prediction plus the background
in the distributions of $E_T^\gamma$, $\Delta R_{\ell\gamma}$, and the
cluster transverse mass defined by
$M_T(\gamma\ell;\nu)=(((m_{\gamma\ell}^2+|{\bf E_T^\gamma}
+{\bf E_T^\ell}|^2)^{\frac{1}{2}}+{\not\!\!E_T})^2-|{\bf E_T^\gamma}+
{\bf E_T^\ell}+{\bf \not\!\!E_T}|^2)^{\frac{1}{2}}.$
Of 25 events CDF observed, 16 events having $M_T(\gamma\ell;\nu)\leq M_W$
are primarily the radiative $W$ decay events plus background.
Similarly, of 23 events D\O\ observed, 11 events are
primarily the radiative $W$ decay events plus background.
The absence of an excess of high $E_T$ photons rules out deviations
from the SM couplings.

To set limits on the anomalous coupling parameters,
a binned maximum likelihood fit was performed on the $E_T^\gamma$ spectrum
for each of the $W(e\nu)\gamma$ and $W(\mu\nu)\gamma$ samples,
by calculating the probability for the sum of the Monte Carlo
prediction and the background to fluctuate to the observed number of events.
The uncertainties in background estimate, efficiencies, acceptance
and integrated luminosity
were convoluted  in the likelihood function with Gaussian distributions.
A dipole form factor with a form factor scale $\Lambda_W=1.5$ TeV
was used in the Monte Carlo event generation.
The limit contours for the CP--conserving
anomalous coupling parameters $\Delta \kappa$ and
$\lambda$ are shown in Fig.~\ref{WGCONT}, assuming that the CP--violating
anomalous coupling parameters $\tilde{\kappa}$ and
$\tilde{\lambda}$ are zero.
For comparison, previous limits obtained by UA2 and CDF from the 1988-89 data
are included.
Current limits on CP--conserving anomalous $WW\gamma$ couplings are:
     \\

${\rm CDF} \ -2.3<\Delta\kappa<2.3\ (\lambda=0),~~~-0.7<\lambda<0.7\
(\Delta\kappa=0),$
   \\

${\rm D\O\ }\ \  -1.6<\Delta\kappa<1.8\ (\lambda=0),~~~-0.6<\lambda<0.6\
(\Delta\kappa=0),$
        \\

\noindent
at the $95\%$ confidence level.
Limits on CP--violating coupling parameters  were within $3-6\%$ of
those obtained for $\Delta\kappa$ and $\lambda.$
It was found that the limits are insensitive to
the form factor for $\Lambda_W>200$ GeV and are well within the constraints
imposed by the $S$-matrix unitarity~\cite{UNIT} with $\Lambda_W=1.5$ TeV.
D\O\  also performed a two dimensional fit including $\Delta R_\ell\gamma$, and
found that the results are within $3\%$ of those obtained from a fit to
the  $E_T^\gamma$ spectrum only.
            \\

\section*{$Z\gamma$ analysis}

The $Z\gamma$ candidates were obtained by searching for events containing
two isolated, high $E_T$, leptons, and an isolated photon.
Table 6 summarizes geometrical and kinematic selection
as well as integrated luminosity used in each channel.
The CDF observed 4 $ee\gamma$ candidates and 4 $\mu\mu\gamma$
candidates~\cite{CDF_ZG}, while the D\O\ observed 4 $ee\gamma$ candidates
and 2 $\mu\mu\gamma$ candidates~\cite{DO_ZG}.
The background estimate, summarized in Table 7, includes
contributions from: $Z+$jets, where a jet is misidentified as a photon;
$Z\gamma$ with $Z\rightarrow \tau\tau.$
Because we require three isolated objects in the final state,
the background in the $Z\gamma$ candidates is small.
The background--subtracted signal agrees well with the SM prediction
calculated using the Monte Carlo program of Baur and Berger.
CDF derived the $Z\gamma$ cross section times $Z\rightarrow \ell\ell$
branching ratio for photons with $\Delta R_{\ell\gamma}>0.7$ and
$E_T^\gamma>7$~GeV
from a combined $e+\mu$ sample:
$\sigma(Z\gamma)\cdot Br(Z\rightarrow\ell\ell)=
5.1\pm 1.9(stat)\pm 0.3(syst)$~pb, in good agreement with the SM prediction  of
$5.2\pm 0.6(stat\oplus syst)$~pb.
Figure~4 and 5 show the data and the SM prediction plus the background
in the distributions of $E_T^\gamma$ and $\ell^+\ell^-\gamma$ invariant mass
for CDF, and $E_T^\gamma$ for D\O\ , respectively.
No significant deviation from  the SM prediction was observed.

\begin{table}
\caption{Summary of $Z\gamma$ event selection.}
\label{table6}
\begin{tabular}{lcccc}
   & \multicolumn{2}{c}{CDF} & \multicolumn{2}{c}{D\O}  \\
\tableline
   & &  & & \\
  & \multicolumn{1}{c}{$ee\gamma$} & \multicolumn{1}{c}{$\mu\mu\gamma$} &
\multicolumn{1}{c}{$ee\gamma$} & \multicolumn{1}{c}{$\mu\mu\gamma$} \\
   & & & & \\
Geometry  & $|\eta_{e1}|<1.1$  & $|\eta_{\mu 1}|<0.6$ &
$|\eta_{e1,2}|<1.1$ & $|\eta_{\mu 1,2}|<1.0$ \\
  & $1.1<|\eta_{e2}|<4.2$ & $|\eta_{\mu 2}|<1.2$ &  $1.5<|\eta_{e1,2}|<2.5$ &
\\
  & & & & \\
 & \multicolumn{2}{c}{$|\eta_\gamma|<1.1$} &
\multicolumn{2}{c}{$|\eta_\gamma|<1.1$,\ $1.5<|\eta_\gamma|<2.5$}\\
  & & & & \\
Kinematics & $E_T^{e1}>20$ & $p_T^{\mu 1,2}>20$ & $E_T^{e1,2}>25$ &
$p_T^{\mu 1}>15$ \\
(in GeV)    & $E_T^{e2}>20,15,10$ &  &  & $p_T^{\mu 2}>8$ \\
   & \multicolumn{2}{c}{$E_T^\gamma>7$} & \multicolumn{2}{c}{$E_T^\gamma>10$}
  \\
   & & & & \\
   & \multicolumn{2}{c}{$\Delta R_{\ell\gamma}>0.7$}
& \multicolumn{2}{c}{$\Delta R_{\ell\gamma}>0.7$}\\
  & & & & \\
$\int Ldt\ {\rm pb}^{-1}$ & $19.7\pm 0.7$ & $18.6\pm 0.7$ & $13.9\pm 1.7$ &
$13.3\pm 1.6$ \\
\end{tabular}
\end{table}
\begin{table}
\caption{Summary of $Z\gamma$ data, backgrounds and the SM predictions.}
\label{table7}
\begin{tabular}{lcccc}
   & \multicolumn{2}{c}{CDF} & \multicolumn{2}{c}{D\O}  \\
\tableline
   & &  & & \\
  & \multicolumn{1}{c}{$ee\gamma$} & \multicolumn{1}{c}{$\mu\mu\gamma$} &
\multicolumn{1}{c}{$ee\gamma$} & \multicolumn{1}{c}{$\mu\mu\gamma$} \\
Source: & & & & \\
\ \ $Z+$jets & $0.4\pm 0.2$ & $0.1\pm 0.1$ & $0.43\pm 0.06$ & $0.02\pm 0.01$ \\
\ \ $Z(\tau\tau)\gamma$ & negligible & negligible & negligible
& $0.03\pm 0.01$ \\
   & & & & \\
Total background & $0.4\pm 0.2$ & $0.1\pm 0.1$ & $ 0.43\pm 0.06$ &
$0.05\pm 0.01$ \\
  & & & & \\
Data & 4 & 4 & 4 & 2 \\
 & & & & \\
Signal & $3.6\pm 2.0$ & $3.9\pm 2.0$ & $3.6^{+3.2}_{-1.9}$ &
$1.95^{+2.6}_{-1.3}$ \\
 & & & & \\
SM prediction & $4.3\pm 0.2$ & $2.8\pm 0.1$ & $3.2\pm 0.5$ & $2.5\pm 0.5$\\
\end{tabular}
\end{table}

Similarly to the $W\gamma$ analysis, limits on anomalous $Z\gamma$ couplings
were obtained by a fit to the  $E_T^\gamma$ spectrum.
Figure 6 shows the current CDF and D\O\ $95\%$ limit contours for
anomalous $ZZ\gamma$ couplings together with the limits from
L3~\cite{L3} experiment and  the constraints from $S$-matrix
unitarity for $\Lambda_Z=500\ {\rm GeV}$.
The pair of $h^Z_{30}$ and $h^Z_{40}$ is CP--conserving, while that of
$h^Z_{10}$ and $h^Z_{20}$ is CP--violating.
Limits on CP--conserving $ZZ\gamma$ couplings are:
     \\

${\rm CDF} \ -3.0<h^Z_{30}<2.9\ (h^Z_{40}=0),~~~-0.7<h^Z_{40}<0.7\
(h^Z_{30}=0),$
   \\

${\rm D\O\ }\  \ -1.9<h^Z_{30}<1.8\ (h^Z_{40}=0),~~~-0.5<h^Z_{40}<0.5\
(h^Z_{30}=0),$
        \\

\noindent
at the $95\%$ confidence level.
Limits on $Z\gamma\gamma$ couplings are the same to within 0.1.
The sensitivity of limits to the form factor scale, $\Lambda_Z$, was studied.
Both CDF and D\O\  data reach the limit set by unitarity for $\Lambda_Z\sim
500\ {\rm GeV}$, which can be interpreted as the sensitivity limit from the
current data.

\section*{Conclusion}

In conclusion, CDF and D\O\ has studied $W\gamma$ and $Z\gamma$ productions
at $\sqrt{s}=1.8$ TeV in electron and muon channels.
The observed photon $E_T$ spectra agree well with the standard model
predictions, yielding limits on anomalous $WW\gamma$, $ZZ\gamma$ and
$Z\gamma \gamma$ couplings.

It is a pleasure to thank the members of the organizing committee,
U.~Baur, S.~Errede and T.~M$\ddot{\rm u}$ller, and the conference staff
for running the conference so smoothly.
I am indebted to my colleagues on D\O\  and the members of CDF electroweak
physics group for their help in preparing the talk.
This work was supported by the Director, Office of Energy Research,
Office of High Energy and Nuclear Physics,
Division of High Energy Physics of the U.S.
Department of Energy under Contract DE-AC03-76SF00098.


\begin{references}

\bibitem{Hagiwara}K.~Hagiwara, R.D.~Peccei, D.~Zeppenfeld and K.~Hikasa,
Nucl.\ Phys.\ {\bf B282}, 253 (1987).
\bibitem{Kim}K.~Kim and Y-S.~Tsai, Phys.\ Rev.\ D {\bf7}, 3710 (1973).
\bibitem{Baur}U.~Baur and E.L.~Berger, Phys.\ Rev.\ D {\bf 41}, 1476 (1990).
\bibitem{CDF}CDF Collaboration, F.~Abe {\it et al.}, Nucl.\ Instrum.\ Methods
{\bf A271}, 387 (1988).
\bibitem{DO}D\O\ Collaboration, S.~Abachi {\it et al.}, Nucl.\ Instrum.\
Methods {\bf A338}, 185 (1994).
\bibitem{CES}CDF Collaboration, F.~Abe {\it et al.}, Fermilab-PUB-94-244-E.
To appear in Phys.\ Rev. D.
\bibitem{MEENA}D\O\ Collaboration, M.~Narain, ``Proceedings of the American
Physical Society Division of Particles and Fileds Meeting," Fermilab (1992),
eds. R.~Raja and J.~Yoh, Vol.2, 1678.
\bibitem{CDF_WG}CDF Collaboration, F.~Abe {\it et al.},
Phys.\ Rev.\ Lett.\ {\bf 74}, 1936 (1995).
\bibitem{DO_WG}D\O\ Collaboration, S.~Abachi {\it et al.},
Fermilab-PUB-95-101-E, Submitted to Phys.\ Rev.\ Lett.
\bibitem{UNIT}U.~Baur and D.~Zeppenfeld, Phys.\ Lett.\ {\bf B201}, 383 (1988).
\bibitem{CDF_ZG}CDF Collaboration, F.~Abe {\it et al.},
Phys.\ Rev.\ Lett.\ {\bf 74}, 1941 (1995).
\bibitem{DO_ZG}G.~Landsberg, these proceedings;
D\O\ Collaboration, S.~Abachi {\it et al.},
Fermilab-PUB-95-042-E. Submitted to Phys.\ Rev.\ Lett.
\bibitem{L3}P.~Mattig, these proceedings; O.~Adrianni {\it et. al},
Phys.\ Lett.\ {\bf B345}, 609 (1995).
\end{references}
\end{document}